\documentclass[letterpaper,twocolumn]{emulateapj} 
\usepackage[german,english,american]{babel} 
\usepackage{graphicx}
\usepackage{natbib}
\usepackage{color}

\def\lesssim{\mathrel{\hbox{\rlap{\hbox{\lower4pt\hbox{$\sim$}}}\hbox{$<$}}}}
\def\gtrsim{\mathrel{\hbox{\rlap{\hbox{\lower4pt\hbox{$\sim$}}}\hbox{$>$}}}}

\begin{document}

\title{The Youngest Known X-ray Binary: Circinus X-1 and Its Natal
  Supernova Remnant}

\author{S. Heinz\altaffilmark{1}}
\email{heinzs@astro.wisc.edu}
\author{P. Sell\altaffilmark{1}}
\author{R.P. Fender\altaffilmark{2,3}}
\author{P.G. Jonker\altaffilmark{4,5,6}}
\author{W.N. Brandt\altaffilmark{7}}
\author{D.E. Calvelo-Santos\altaffilmark{3}}
\author{A.K. Tzioumis\altaffilmark{8}}
\author{M.A. Nowak\altaffilmark{9}}
\author{N.S. Schulz\altaffilmark{9}}
\author{R. Wijnands\altaffilmark{10}}
\author{M. van der Klis\altaffilmark{10}}

\altaffiltext{1}{Department of Astronomy, University of Wisconsin-Madison,
  Madison, WI 53706, USA} 
\altaffiltext{2}{University of Oxford, Astrophysics, Oxford OX1 3RH,
  UK}
\altaffiltext{3}{Physics and Astronomy, University of Southampton,
  Highfield, Southampton SO17 1BJ, UK}
\altaffiltext{4}{SRON, Netherlands Institute for Space Research, 3584
  CA, Utrecht, the Netherlands}
\altaffiltext{5}{Department of Astrophysics/IMAPP, Radboud University
  Nijmegen, 6500 GL, Nijmegen, The Netherlands}
\altaffiltext{6}{Harvard--Smithsonian Center for Astrophysics,
  Cambridge, MA~02138, USA}
\altaffiltext{7}{Department of Astronomy \& Astrophysics, The
  Pennsylvania State University, University Park, PA 16802, USA}
\altaffiltext{8}{Australia Telescope National Facility, CSIRO, Epping,
  NSW 1710, Australia}
\altaffiltext{9}{Kavli Institute for Astrophysics and Space Research,
  Massachusetts Institute of Technology, Cambridge, MA 02139, USA}
\altaffiltext{10}{Astronomical Institute ``Anton Pannekoek'', University
  of Amsterdam, 1090 GE Amsterdam, The Netherlands}

\begin{abstract} Because supernova remnants are short lived, studies
  of neutron star X-ray binaries within supernova remnants probe the
  earliest stages in the life of accreting neutron stars.  However,
  such objects are exceedingly rare: none were known to exist in our
  Galaxy.  We report the discovery of the natal supernova remnant of
  the accreting neutron star Circinus X-1, which places an upper limit
  of $t<4,600$ years on its age, making it the youngest known X-ray
  binary and a unique tool to study accretion, neutron star evolution,
  and core collapse supernovae.  This discovery is based on a deep
  2009 {\em Chandra} X-ray observation and new radio observations of
  Circinus X-1. Circinus X-1 produces type I X-ray bursts on the
  surface of the neutron star, indicating that the magnetic field of
  the neutron star is small.  Thus, the young age implies either that
  neutron stars can be born with low magnetic fields or that they can
  rapidly become de-magnetized by accretion.  Circinus X-1 is a
  microquasar, creating relativistic jets which were thought to power
  the arcminute scale radio nebula surrounding the source.  Instead,
  this nebula can now be attributed to non-thermal synchrotron
  emission from the forward shock of the supernova remnant.  The young
  age is consistent with the observed rapid orbital evolution and the
  highly eccentric orbit of the system and offers the chance to test
  the physics of post-supernova orbital evolution in X-ray binaries in
  detail for the first time.
\end{abstract}


\section{Introduction}

Observations of young high-mass X-ray binary (HMXB) neutron stars can
provide powerful constraints on the physics of accretion, supernova
explosions and the early evolution and birth properties of neutron
stars.  Important properties that can be constrained from observations
of such systems include the age, the magnetic field strength, and the
spin of a newly born neutron star, the mass of the neutron star, the
mass of its progenitor star, the mass of its companion star, supernova
kick velocities, and the pre- and post-supernova orbital parameters of
the binary system and their evolution.

Identifying newly formed HMXBs requires the detection of the supernova
remnant in which they formed, which allows an accurate determination
of their age.  This motivates the search for supernova remnants around
Galactic HMXBs.  However, the time during which the remnant is visible
is at least three orders of magnitude shorter than typical binary and
stellar evolution time scales, making X-ray binaries within their
remnants extremely rare objects.  The only firmly established X-ray
binary within a supernova remnant in our Galaxy, the X-ray binary
SS433 \citep{geldzahler:80}, most likely does not contain a neutron
star but a black hole.

The neutron star\footnote{The unambiguous identification of Circinus
  X-1 as a neutron star is based on the detection of type I X-ray
  bursts from the source \citep{tennant:86,linares:10}} X-ray binary
Circinus X-1 has historically been among the brightest X-ray sources
in the sky.  The often erratic accretion behavior, the unusually
powerful jets of this microquasar, the difficulty in identifying the
mass and type of the companion star, and the large scale radio nebula
historically thought to be produced by the jets
\citep{stewart:93,tudose:06} have made it hard to classify the source
in the context of typical neutron star X-ray binary classification
schemes \citep[e.g.][]{oosterbroek:95,calvelo:12b}.  The source has
often been referred to as a low-mass X-ray binary, which would imply
an old age for Circinus X-1, but many of its characteristics suggest
that the source might, in fact, be very young
\citep{clarkson:04,jonker:07}.

Circinus X-1's large X-ray flux has made searches for faint diffuse
X-ray emission around the source all but impossible, because the X-ray
light from the neutron star is scattered into an arcminute scale halo
by interstellar dust, easily overwhelming any underlying low surface
brightness emission.  In this paper, we present the discovery of the
supernova remnant of the neutron star X-ray binary Circinus X-1.

Literature estimates for the distance to Circinus X-1 range from
$4\,{\rm kpc}$ \citep{iaria:05} to $11\,{\rm kpc}$, with a most likely
value of $D=8-10.5\,{\rm kpc}$ \citep[based on both the
radius-expansion burst method and the observed Galactic neutral
Hydrogen column density,][]{jonker:04}.  Throughout this paper, we
will use a fiducial distance of $D=8\,{d_{8}}\,{\rm kpc}$
to the source, with explicit dependence of all numerical values on the
true distance expressed in terms of $d_{8}$.  The qualitative results
presented in this paper are insensitive to the actual value of
$d_{8}$.

\section{Observations}
\subsection{Data Reduction and Image Analysis}

Circinus X-1 entered an extended period of low flux in about 2005,
which allowed the detection of extended X-ray synchrotron emission
from the powerful jets of this microquasar \citep{heinz:07,soleri:09}.
We re-observed the source for 98,723 seconds with the Advanced CCD
Imaging Spectrometer on {\em Chandra} on May 1, 2009\footnote{Obs ID
  10062 in the {\em Chandra} archive} to confirm this detection and
study the jet emission while the source was at the lowest flux state
ever recorded.  Data were taken in timed exposure mode and telemetered
in Faint mode. The data were reprocessed, reduced, and analyzed using
CIAO version 4.5, CALDB version 4.5.6, and {\tt XSPEC} version 12.8.0.

An analysis of the jet emission was presented in a previous paper
\citep{sell:10}.  This analysis did not address the large scale
diffuse X-ray emission beyond the synchrotron jets in detail, which
was tentatively interpreted as emission from the dust scattering halo.
Below, we present a full investigation of the arcminute scale diffuse
emission, showing clearly that the emission is, in fact, not due to
dust scattering.

\begin{figure}
  \center\resizebox{\columnwidth}{!}{\includegraphics{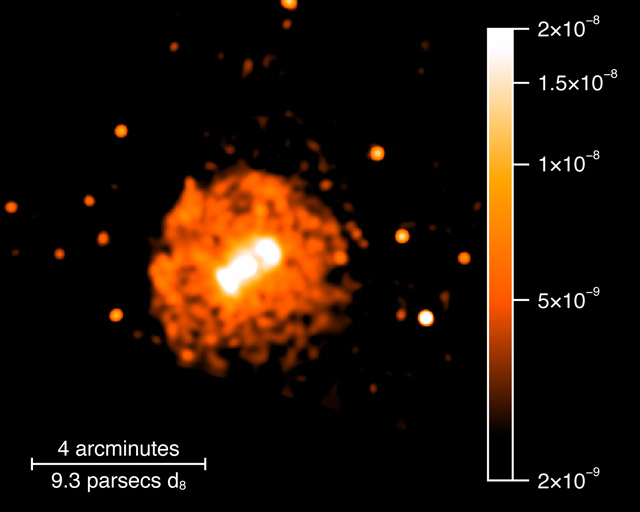}}
  \caption{{\em\,Chandra} X-ray image of the supernova remnant and jet
    of Circinus X-1 in the energy range 1-3 keV.  Adaptive binning to
    a signal-to-noise of 4 per resolution element and Gaussian
    smoothing with 10 pixel FWHM were applied for noise reduction.
    The CCD read-out streak was removed from the image according to
    standard practices.  The central elongated brightness enhancement
    (yellow-white color) is due to the neutron star point source and
    the bi-polar jet emission \citep{heinz:07,soleri:09,sell:10}.  The
    scale bar shows an angular scale of 4 arcminutes.  The color bar
    shows the surface brightness scale in units of
    photons\,s$^{-1}$\,cm$^{-2}$\,arcsec$^{-2}$.  The image is
    oriented such that North is up and East is to the Left.}
  \label{fig:smoothed}
\end{figure}

Careful inspection of the faint, diffuse emission in observation Obs
ID 10062 beyond the synchrotron jets reveals a smooth, roughly
circular X-ray nebula, shown in Fig.~\ref{fig:smoothed}, that extends
about 2.5 arcminutes from the point source, beyond which the emission
shows a well defined edge in the Northern half of the image (the
Southern half is not fully covered by the CCDs).  In order to
visualize the low surface-brightness excess, we applied adaptive
binning to the image, using weighted Voronoi-Delaunay tessellation
with a minimum signal-to-noise of 4 per resolution element \citep[wvt
image, see][]{diehl:06} and smoothed the resulting image using a
Gaussian with a FWHM of 10 {\em Chandra ACIS} pixels.

\subsection{Eliminating Dust Scattering}

Dust scattering of emission from the bright, central accreting neutron
star source could in principle produce diffuse emission on the scales
of the observed nebula.  Dust scattering can be ruled out as the
dominant source of the extended X-ray emission in Obs ID 10062 on
three grounds: (a) Given historical observations of the dust
scattering halo of Circinus X-1, the X-ray nebula is too bright to be
due to dust scattering, (b) the X-ray spectrum of the nebula is
inconsistent with the expected powerlaw emission of the dust
scattering halo, and (c) a chance coincidence of the close
morphological correspondence between the X-ray and radio nebulae is
highly unlikely.  We will discuss these arguments in turn.

\subsubsection{Surface Brightness Arguments Against 
    Dust Scattering}
The central brightness enhancement in Fig.~\ref{fig:smoothed} is due
to the neutron star point source and the bi-polar jet emission.
However, the emission flattens at radii between 30 and 150 arcseconds,
deviating from the expected, gradually declining, radial brightness
profile of dust scattering.

\begin{figure}
  \center\resizebox{\columnwidth}{!}{\includegraphics{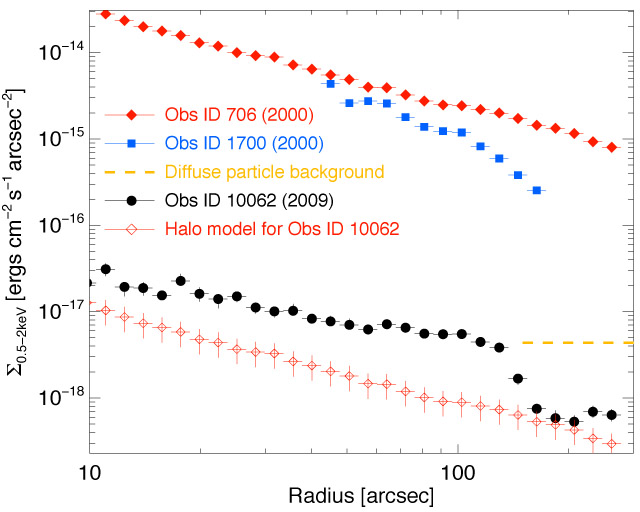}}
  \caption{The 0.5-2 keV X-ray surface brightness $\Sigma$ of {\em
      Chandra} observations Obs ID 706, 1700, and 10062 shown as solid
    red diamonds, blue squares, and black dots, respectively, plotted
    against radial distance from Circinus X-1.  Diffuse particle
    background emission at a uniform level of $\Sigma_{0.5-2.0} \sim
    4.4\times 10^{-18}\,{\rm ergs\,s^{-1}\,cm^{-2}\,arcsec^{-2}}$ was
    subtracted from the Obs ID 10062 profile (shown as dashed yellow
    line for comparison). Regions dominated by synchrotron jet
    emission were excluded.  The estimated contribution from the dust
    scattering halo to Obs ID 10062 is shown as open red diamonds,
    well below the measured surface brightness of the supernova
    remnant.  Vertical error bars indicate one-sigma uncertainties.}
  \label{fig:profiles}
\end{figure}

In order to eliminate dust scattering as the source of the extended
emission, we constructed radial surface brightness profiles $\Sigma$
of the dust scattering halo from two previous {\em Chandra}
observations (Obs ID 706, March 2000; Obs OD 1700, June 2000), shown
in Fig.~\ref{fig:profiles}.  During these observations, the X-ray
binary point source was a factor of approximately 2,500
and 1,000 brighter than during Obs ID 10062, respectively.  Given
these high fluxes and the existing estimate of the dust scattering
fraction for Circinus X-1 \citep{predehl:95}, any extended emission
beyond the inner few arcseconds in Obs ID 706 and Obs ID 1700 must be
due to dust scattering.

Dust scattering induces a time delay of $\Delta t =
  1.1\,d_{\rm 8}\,{\rm days}(\theta/100'')^2x/(1-x)$ between the
arrival of light from the point source and the arrival of the
corresponding scattered dust emission \citep{predehl:96} where
$\theta$ is the angle of the scattering halo on the sky and $x$ is the
fractional distance from the observer to the dust.  We must therefore
consider the point source emission not just during the observation but
also during the days leading up to each observation.

\begin{figure}
  \center\resizebox{\columnwidth}{!}{\includegraphics{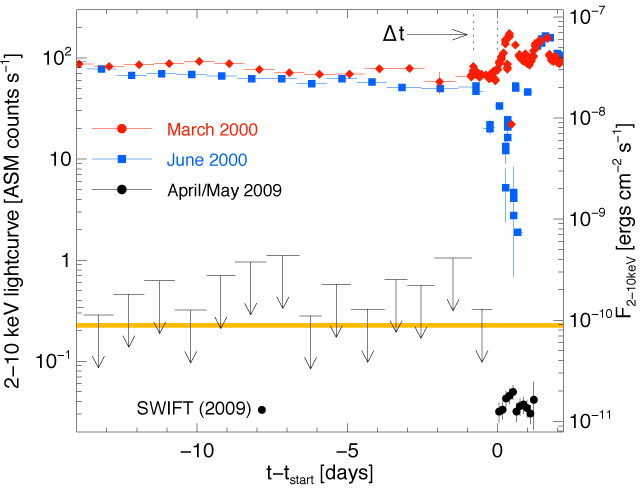}}
  \caption{X-ray lightcurves of Circinus X-1.  {\em Rossi X-ray Timing
      Explorer All Sky Monitor} lightcurves of the point source
    emission from Circinus X-1 during the time interval $t_{\rm start}
    - 14\,{\rm days}$ to $t_{\rm start} + 2\,{\rm days}$, measured
    from the start times $t_{\rm start}$ of Obs IDs 706, 1700, and
    10062 (red diamonds, blue squares, and black upper limits,
    respectively).  The {\em Chandra} and {\em Swift} fluxes during
    and before Obs ID 10062 are shown as black dots.  The inferred
    dust-delay time for the outer halo of $\Delta t \lesssim 1\,{\rm
      day}$ is shown by the horizontal arrow.  The yellow horizontal
    bar indicates the point source flux required to produce a halo as
    bright as the nebula in Obs ID 10062.  Vertical error bars and
    upper limits indicate one-sigma uncertainties.  There is no
    indication from the observations that the source could have been
    bright enough just prior to {\em Chandra} Obs ID 10062 to produce
    the observed nebula by dust scattering.}
  \label{fig:lightcurves}
\end{figure}

The X-ray lightcurves for the two weeks prior to each of the three
observations are shown in Fig.~\ref{fig:lightcurves}.  The daily
average {\em Rossi X-ray Timing Explorer All Sky Monitor (ASM)} flux
for Circinus X-1 in the two weeks prior to Obs ID 706 was
approximately constant at $\approx$80 counts per second.  The analysis
of the point source and the dust scattering halo was restricted to the
first 40\% of Obs ID 706, representative of the flux the source had in
the two weeks leading up to Obs ID 706, avoiding the bright flare
during the second half of the observation visible in
Fig.~\ref{fig:lightcurves}.

From the dust scattering emission in Obs ID 706, we constructed the
expected radial surface brightness profile of the dust scattering
contribution to Obs ID 10062, using the ratio of the
  0.5-2.0 keV point source fluxes of 2720 measured with {\em Chandra}
for both observations (open red diamonds in Fig.~\ref{fig:profiles};
error bars account for the variability of the source in the week
leading up to Obs ID 706, estimated from the standard deviation of the
orbit-to-orbit ASM flux variations, propagated in quadrature with
the uncertainties in the radial surface brightness
  profile of Obs ID 706 and the point source flux of Obs ID 10062.

The radial surface brightness profile for Obs ID 10062 itself,
excluding the regions of the image identified as synchrotron shock
emission \citep{sell:10}, is shown as black dots in
Fig.~\ref{fig:profiles}.  Over the extent of the nebula, the observed
surface brightness in the 0.5-2 keV energy band is a factor of
approximately 3 to 6 higher than the predicted contribution from the
dust scattering halo.  Thus, the nebula cannot have been produced by
dust scattering while the point source flux was at the level measured
{\em during} our {\em Chandra} observation.

Using Obs ID 1700, we can estimate the delay time $\Delta t$ for the
dust towards Circinus X-1 to show the nebula in Obs ID 10062 cannot
have been produced by a flare of the point source {\em prior} to Obs
ID 10062: In the two weeks prior to Obs ID 1700, the {\em ASM} X-ray
lightcurve of Circinus X-1 was roughly constant at 50-70 counts per
second, close the flux before Obs ID 706.  One day before Obs ID 1700,
the {\em ASM} source flux dropped sharply to 10-20 counts/s and then
re-brightened to approximately 30 counts per second during the {\em
  Chandra} observation.

The halo surface brightness within the inner 100 arseconds of Obs ID
1700 is at roughly 50\% of the halo surface brightness in Obs ID 706,
approximately the same ratio as the point source fluxes during the
observations.  Outward of 100 arcseconds, the halo brightness drops
steeply in Obs ID 1700 to about 10\%-20\% of the surface brightness in
the outer halo of Obs ID 706.  In the two week intervals leading up to
both observations, the {\em only} time period before Obs ID 1700 when
the point source was a factor of 5 to 10 dimmer than it was at the
corresponding time before Obs ID 706 was during the 24 hours leading
up to the observations.  Thus, the dust delay time for annuli between
100 and 170 arcseconds from the central source must be $\Delta t \leq
1\,{\rm day}$, and correspondingly shorter on smaller angular scales.

During the second half of Obs ID 706, the dust scattering halo shows a
$\sim 25\%$ increase in flux, roughly within 10,000 seconds of the
point source brightening.  This yields an independent estimate of
$\Delta t \sim 0.4\,{\rm days}$ for the variability and thus the dust
delay time of the halo, consistent with the estimate of $\Delta t \leq
1\,{\rm day}$ from Obs ID 1700.

On the other hand, the X-ray nebula in Obs ID 10062 shows no sign of
X-ray variability in any of the annuli between 20 and 150 arcseconds,
consistent with the expectation for diffuse emission from an
astrophysical source like a supernova remnant, but inconsistent with
dust scattering from a bright flare of the point source during the
time leading up to the observation.  We can place an upper limit on
any secular change of the surface brightness during the observation of
less than $2.5\%$ per day.

Obs ID 10062 had an exposure time longer than 1 day, and the point
source flux during that observation was constant within the
uncertainties.

\subsubsection{Spectral Evidence Against Dust Scattering}
The X-ray spectrum of the nebula shows clear evidence for emission
lines from Magnesium, Silicon, and Sulfur (see
  \S\ref{sec:fits} for a detailed discussion of the spectrum).  On
the other hand, the point source spectrum is best fit by a powerlaw
$F_{\nu} \propto \nu^{0.4}$, modified by photoelectric absorption.
Dust scattering is expected to steepen this powerlaw by a factor of
$\nu^{-2}$.  The resulting powerlaw spectrum is statistically
inconsistent with the observed thermal nebular spectrum.  It is
consistent with the residual background emission in the annuli beyond
the nebula, indicating that residual dust scattering emission is
present at the low level predicted by the profile derived from Obs ID
706.

\subsubsection{Morphological Evidence Against Dust Scattering}
Finally, and most importantly, the sharp edge of the northern half of
the X-ray nebula closely follows the edge of the radio nebula
(Fig.~\ref{fig:color}), including several deviations from
the generally circular shape of the northern hemisphere and the
east-west asymmetry that would not be expected for an axi-symmetric
dust scattering halo.  It is highly unlikely that two independent
mechanisms (dust scattering and the shock responsible for the radio
emission) would produce such a close match between these two shapes.

We conclude that the bulk of the extended emission in Obs ID 10062
cannot be caused by dust scattering.

\subsection{ATCA Radio Observations}

We obtained a high-resolution radio image of the remnant in the
1.1-3.1 GHz band, taken with the Australia Telescope Compact Array on
December 16, 2011, with an integration time of 19.5 hours in 6A array
configuration (minimum baseline of 337m, maximum of 5939m).  The FWHM
beam size of the image is 4.9 $\times$ 4.0 arcsec at position angle
0.5$^{\circ}$, with a theoretical rms noise of 4$\mu{\rm Jy}$ per
beam.  PKS1511-55 was used for flux, phase, and band-pass calibration.

Deconvolution was carried out using a combination of the {\tt MFCLEAN}
\citep{sault:94} and {\tt CLEAN} \citep{hogbom:74} subroutines.  Data
processing was carried out in {\tt MIRIAD} \citep{sault:95}.

Previous radio observations of the nebula were taken at lower
resolution \citep{stewart:93,tudose:06}, suffering from contamination
of the large scale diffuse emission by the bright jet and point source
emission.  Higher resolution images of Circinus X-1, taken to
investigate the sub-arcsecond jet emission close to the point source
typically over-resolve the diffuse emission, but recent 5.5 GHz
observations show evidence for a sharp, filamentary edge to the nebula
\citep{calvelo:12b}.

\begin{figure}
  \center\resizebox{\columnwidth}{!}{\includegraphics{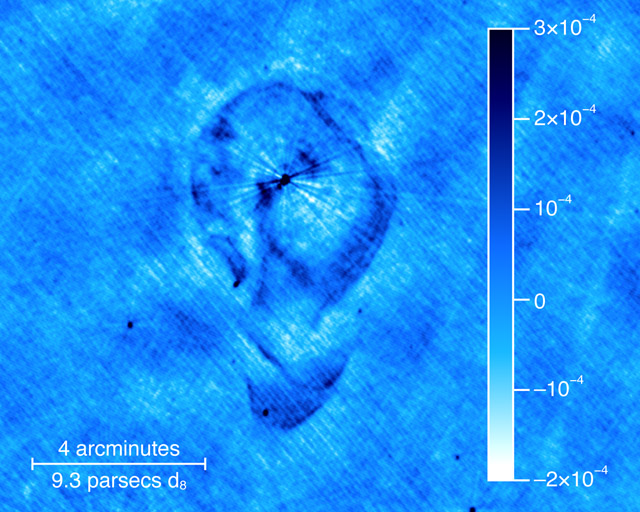}}
  \caption{1.1-3.1 GHz broad band radio image of the Circinus X-1
    supernova remnant and jet taken with the Australia Compact
    Telescope Array on 16 December 2011.  The scale bar shows an
    angular scale of 4 arcminutes.  The color bar shows the image
    surface brightness in units of Jy per beam, with a FWHM beam size
    of 4.9$\times$4.0 arcseconds at position angle 0.5$^{\circ}$.  The
    image is oriented such that North is up and East is to the Left.}
  \label{fig:radio}
\end{figure}

\begin{figure}
  \center\resizebox{\columnwidth}{!}{\includegraphics{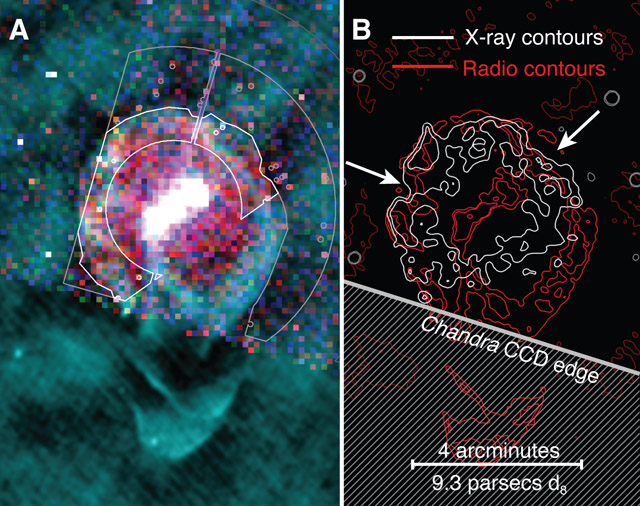}}
  \caption{Radio-X-Ray Overlay of the Circinus X-1 Supernova
    Remnant. {\em Left panel A:} soft 1-2 keV, medium 2-3 keV, and
    hard 3-5 keV X-ray emission in red, green, and blue, respectively;
    the radio synchrotron emission is overlaid in Cyan.  The
    extraction regions for the source spectrum and the inner
    background spectrum are indicated as white and light-gray
    contours, respectively.  The inner rim of the outer background
    extraction region can be seen as the dark-gray contour.  {\em
      Right panel B:} overlay of the radio (red) and X-ray (white)
    contours of the remnant edge.  Contours we identify as part of the
    remnant are shown in a lighter shade.  The part of the image not
    covered by the {\em\,Chandra} CCDs is marked by the grey hatched
    area. The white arrows indicate dents where both radio and X-ray
    emission deviate from a simple circular contour.  The images are
    oriented such that North is up and East is to the Left.}
  \label{fig:color}
\end{figure}

The radio image presented here clearly resolves the nebula as an
edge-brightened, asymmetric synchrotron shell, shown in
Fig.~\ref{fig:radio}.  Figure \ref{fig:color} shows an overlay of this
radio image on the diffuse X-ray emission.  The portions of the nebula
covered by the {\em Chandra} CCDs line up well with the radio nebula:
The edge of the X-ray emission is just inside the radio shell, even
following a pronounced dent in the North-Western section of the radio
image.  The morphology of both the X-ray and the radio images are
consistent with the expected appearance of a young supernova remnant.

\subsection{Extraction of the Remnant X-ray Spectrum:}

Because of the potential contamination by residual emission from the
dust scattering halo, we used only the outer portion of the remnant
beyond 95 arcseconds from the source, where the fractional
contribution from dust scattering should be smallest, to determine the
remnant spectrum.  We restricted the spectral extraction to the
northern portion of the nebula, where the surface brightness of the
nebula and therefore the signal-to-noise is largest.  This section of
the nebula is fully covered by the {\em ACIS S-3 CCD}.  Point sources,
the read-out streak, and regions affected by the dithered edge of the
{\em ACIS} S-3 chip were excluded from the extraction.

\begin{figure}
  \center\resizebox{\columnwidth}{!}{\includegraphics{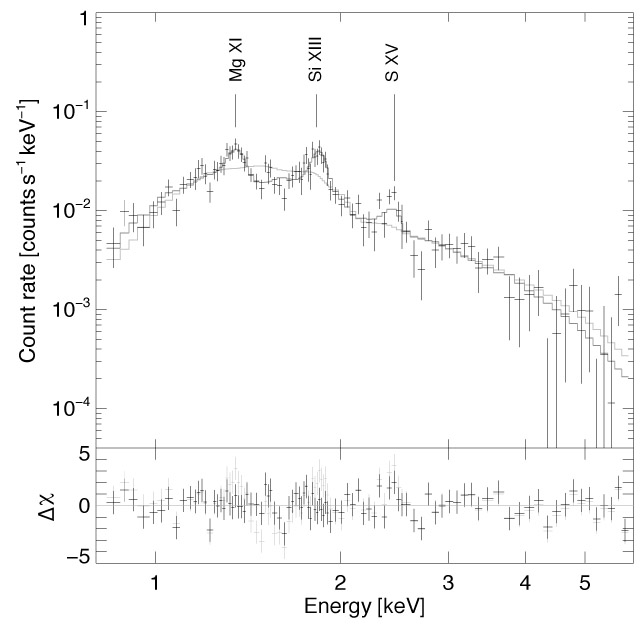}}
  \caption{{\em Chandra} {\em Top panel:} X-ray Spectrum of the outer
    portion of the supernova remnant, showing Magnesium XI and XII
    (1.4 keV), Silicon XIII (1.8 keV) and Sulfur XV (2.4 keV) emission
    lines.  Error bars indicate one-sigma uncertainties.  The dark
    gray histogram shows the best fit {\tt SEDOV} shock model with
    photoelectric foreground absorption, the light-grey histogram
    shows the best-fit powerlaw model.  The spectrum contains
    approximately 2900 net source counts and covers approximately 30\%
    of the remnant area visible on the ACIS CCDs.  {\em Bottom panel:}
    fit residuals for the best fit {\tt SEDOV} and powerlaw models in
    dark and light grey, respectively.}
  \label{fig:spectrum}
\end{figure}

We chose the primary background region to be close to the remnant in
order to model potential residual emission from the dust scattering
halo.  A second background region on the farthest corner of the same
{\em Chandra ACIS} S-3 CCD was chosen to subtract the particle
background.  The resulting spectrum contains approximately 2900 net
source counts in the 0.8-6 keV range used for the fits, out of 7774
total (source plus background) counts.  The background-subtracted
spectrum used in the fits is shown in Fig.~\ref{fig:spectrum}.  The
source and inner background extraction regions are indicated in
Fig.~\ref{fig:color}.

From the radial surface brightness profile of the halo, we estimate
the maximum contamination by dust scattering to be $< 20\%$ of the
0.5-3 keV flux from the remnant.  We modeled this contribution by
jointly fitting a powerlaw modified by photoelectric absorption to
both the remnant and the primary background region, in addition to the
Sedov blast wave model, which we fitted to the remnant spectrum only.

The relative flux ratio of the halo emission contaminating the
spectrum of the remnant to the halo emission in the primary background
spectrum was derived from the radial surface brightness profile of the
dust scattering halo in Obs ID 706 derived in the appendix and shown
in Fig.~\ref{fig:profiles}, integrated over the regions from which the
spectra were extracted.  Particle background contamination (subtracted
from the spectrum using the second background region) dominates the
spectrum above energies of 6 keV and below energies of 0.8 keV, which
motivated the choice of the energy band of 0.8 to 6 keV used in the
fits in \S\ref{sec:fits}.

\subsection{Spectral Fits}
\label{sec:fits}

The X-ray spectrum of the nebula, shown in Fig.~\ref{fig:spectrum},
clearly indicates significant photoelectric attenuation by foreground
gas and dust.  The interstellar neutral hydrogen column density of
$N_{\rm H} \approx 1.8 \pm 0.1\times 10^{22}\,{\rm cm^{-2}}$
determined from the spectral fits\footnote{Uncertainties are quoted at
  the one-sigma 68\% confidence level unless otherwise specified} is
consistent with a source distance of $D \approx 8\,{\rm
    kpc}$ \citep{jonker:04}.  The large column density is consistent
with previous estimates and the large amount of visible extinction
towards Circinus X-1.

The X-ray spectrum shows clear evidence for emission lines from
Magnesium, Silicon, and Sulfur, inconsistent with a simple powerlaw
model, which would be expected if the X-rays were due to synchrotron
emission.  Figure \ref{fig:spectrum} shows the best fit powerlaw model
and residuals in the top and bottom panels, respectively.  With a
reduced chi-square of $\chi_{\rm red, power}^2 = \frac{197.63}{91{\rm
    d.o.f.}} = 2.2$, we can formally rule out a powerlaw model for the
X-ray spectrum with a high degree of confidence.  Adding narrow
Gaussian emission lines for Magnesium, Silicon, and Sulfur improves
the chi-square by $\Delta \chi^2_{\rm Mg} = 30.27$, $\Delta
\chi^2_{\rm Si} = 36.31$, and $\Delta \chi^2_{\rm S} = 24.58$,
respectively.

On the other hand, non-equilibrium shock models such as {\tt PSHOCK,
  NEI}, and {\tt SEDOV} provide a statistically satisfactory spectral
fit, with a reduced chi-square of $\chi^2_{\rm red, non-eq} = 1.0$
in all three cases, consistent with the expectations for
a young supernova remnant.  Non-equilibrium models are statistically
preferred over equilibrium ionization models like {\tt APEC}, which
give a best-fit reduced chi-square of $\chi^2_{\rm red, eq} = 1.2$.

\begin{figure}
  \center\resizebox{\columnwidth}{!}{\includegraphics{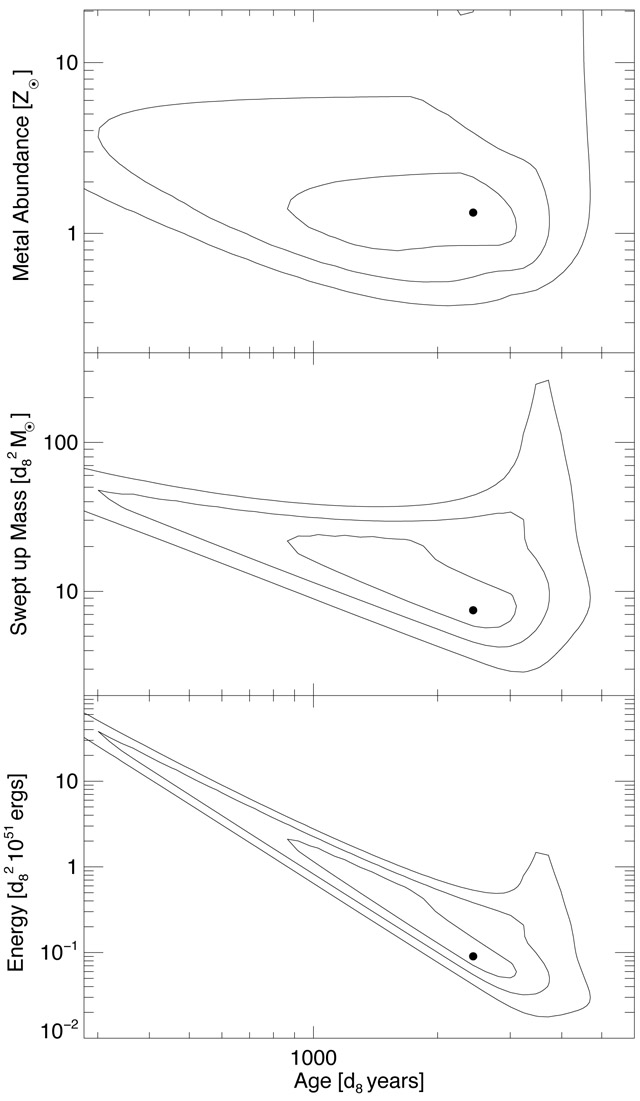}}
  \caption{Confidence intervals from the {\tt SEDOV} model fits for
    the remnant age in years (horizontal axis), the supernova energy
    in units of $10^{51}\,{\rm ergs}$, the swept up mass in units of
    solar masses, and the metal abundance relative to solar values
    (bottom, middle, and top panel, respectively).  From inside out,
    the contours show the one (68\%), two (95\%), and three-sigma
    (99.7\%) confidence ranges. The fits provide robust upper limits
    on the remnant age and lower limits on the swept up mass and
    supernova energy.  The formal best fit parameters are marked by
    black dots.}
  \label{fig:contours}
\end{figure}

Based on the evidence from both imaging and spectral analysis, we
conclude that the X-ray nebula must be the supernova remnant of
Circinus X-1.  We estimate the supernova parameters by fitting the
spectrum with a {\tt SEDOV} blast wave model
\citep{hamilton:83,borkowski:01}.  The best fit model is shown as in
black in Fig.~\ref{fig:spectrum}.  Fit results are listed in Table
\ref{tab:fit} and confidence intervals of the key supernova parameters
derived from the {\tt SEDOV} model fits are plotted in
Fig.~\ref{fig:contours}.

\begin{deluxetable}{cccccc}
  \tabletypesize{\scriptsize} 
  \tablecaption{{\tt PHABS*SEDOV} best fit
    parameters\tablenotemark{a}.\label{tab:fit}}

  \tablehead{\colhead{$N_{\rm H,22}$\tablenotemark{b}}& 
    \colhead{$T_{\rm e}$\tablenotemark{c}}&
    \colhead{$T_{\rm s}$\tablenotemark{d}}& 
    \colhead{$Z/Z_{\odot}$\tablenotemark{e}}&
    \colhead{$\tau_{11}$\tablenotemark{f}}& 
    \colhead{EM$_{\rm L,17}$\tablenotemark{g}}}  
  \startdata $1.8\pm 0.1$ &
  $0.9^{+0.4}_{-0.8}$ & $1.0^{7.3}_{-0.4}$ & $1.3^{+0.9}_{-0.5}$ &
  $1^{+1}_{-0.2}$ & $1.5^{+1.3}_{-0.7}$
  \enddata
  
  \tablenotetext{a}{The reduced chi-square of the best fit parameters
    shown in the table is $\chi^{2}/{\rm d.o.f.} = 348.5/345$.}
  \tablenotetext{b}{Absorption column density $N_{\rm H,22}\equiv
    N/10^{22}\,{\rm cm^{-2}}$ in units of $10^{22}\,{\rm cm^{-2}}$}
  \tablenotetext{c}{Electron temperature $T_{\rm e}$ in keV}
  \tablenotetext{d}{Shock temperature $T_{\rm s}$ in keV}
  \tablenotetext{e}{Metal abundance $Z$ in units of solar metalicity
    $Z_{\odot}$} \tablenotetext{f}{Ionization age $\tau_{11}\equiv
    \tau/10^{11}\,{\rm s}$ in units of $10^{11}\,{\rm s}$, where
    $\tau=n_{e}t$ is the product of the electron density $n_{e}$
    behind the shock and the age of the remnant $t$}
  \tablenotetext{g}{Spectral normalization, given as the line emission
    measure $EM_{\rm L} = \int dl n_{\rm e}n_{\rm H}$, averaged over
    the spectral extraction region, in units of $10^{17}\,{\rm
      cm^{-5}}$}
\end{deluxetable}

The formal best fit shock temperature is $T_{\rm
  s}=1.0^{+7.3}_{-0.4}\,{\rm keV}$, corresponding to a shock velocity
of $v_{\rm s}=910^{+1690}_{-210}\,{\rm km\,s^{-1}}$ and a remnant age
of $t=2440^{+720}_{-1590}\,d_{8}\,{\rm yrs}$.  The
dynamical remnant parameters, such as the age, the swept up mass, and
the explosion energy, are derived from the spectral fits using the
standard relations for adiabatic supernova remnants in the Sedov
expansion phase \citep[][equations 4a-4e; see also, e.g.,
\citealt{safi-harb:00,borkowski:01}]{hamilton:83}.

The {\em upper} limit of the shock temperature and the lower limit to
the remnant age are not well constrained by the {\tt SEDOV} fits,
consistent with the expectations for young supernova remnants
\citep{borkowski:01,vink:12}.  This is due to (a) the uncertainty
introduced by the residual contribution from the dust scattering
spectrum, (b) the dominant contribution of the diffuse particle
background above energies of 6 keV, and (c) the low effective area of
the {\em Chandra} mirrors above above 3 keV, with very few sources
counts in the 3-6 keV range.  These sources of uncertainty are fully
reflected in the confidence intervals plotted in
Fig.~\ref{fig:contours} and quoted in the text.

More importantly, however, the fit provides a robust three-sigma {\em
  lower} limit of $T_{\rm s} > 0.3\,{\rm keV}$ on the shock
temperature.  Given the nominal source distance of
$D=8\,d_{8}\,{\rm kpc}$ and a measured shock radius of $R
\sim 150$ arcsec in the northern hemisphere of the remnant, this lower
limit on the shock temperature translates into a three sigma upper
limit of $t < 4,600\,d_{8}\,{\rm yrs}$ on the age of the
remnant.  This result is insensitive to the specific shock model used:
fits with {\tt SEDOV, NEI, PSHOCK, NPSHOCK} or even equilibrium ({\tt
  APEC}) models yield the same limit.

We determine the mean ambient Hydrogen density $n_{0}$ from the
ionization age parameter $\tau \equiv n_{\rm e} t = 4.8n_{0} t$
\citep[for cosmic abundance plasma and standard strong-shock
Rankine-Hugoniot jump conditions for a $\gamma=5/3$ ideal gas, where
$n_{\rm e}$ is the post-shock electron density][]{safi-harb:00} to be
$n_{0} \approx 0.27^{+0.60}_{-0.05}\,d_{8}^{-1}\,{\rm
    cm^{-3}}$.  The low external density estimate from the ionization
age is roughly consistent with estimates derived from the emission
measure, $n_{0,EM} \approx
  0.16^{+0.22}_{-0.11}\,d_{8}^{-1/2}\,f^{-1/2}\,{\rm cm^{-3}}$, where
$f \leq 1$ is the filling factor of the emitting gas.

The Sedov estimate of the total energy in the blast wave is
$E_{\rm s} \sim 0.09^{+2.2}_{-0.03}\times
  10^{51}\,d_{8}^{2}\,{\rm ergs}$.  Within the uncertainties, this is
consistent with typical supernova remnant energies of $E \sim
10^{51}\,{\rm ergs}$.

From the spectral fits, we find that the abundance of heavy elements
in the outer remnant is $Z = 1.3^{+0.9}_{-0.5} Z_{\odot}$ relative to
solar, which is consistent with emission predominantly from the
forward shock into relatively un-enriched material.  Because of the
possible contamination of the inner remnant by dust scattering
emission, we cannot reliably test for enriched emission by the
supernova ejecta.

\section{Discussion and Conclusions}

The upper limit of $t < 4,600\,d_{8}\,{\rm years}$ on the
age of the remnant makes Circinus X-1 the youngest known X-ray binary.
The black hole candidate SS433, the only other firmly
established\footnote{The nature of the central source in the young
  supernova remnant RCW103 is currently uncertain \citep{li:07}, but
  it could be a low-mass X-ray binary at an age similar to Circinus
  X-1 \citep{deluca:06}.}  Galactic X-ray binary within a supernova
remnant \citep{geldzahler:80}, has an estimated age of $10^{4}\,{\rm
  yrs}< t \lesssim 10^{5}\,{\rm yrs}$ \citep{lockman:07,goodall:11}.
Two other plausible HMXB candidates in nearby galaxies have been
suggested to reside within supernova remnants, the Be/X-ray binary
pulsar SXP 1062 \citep{henault-brunet:12} in the Small Magellanic
Cloud, and DEM L241 \citep{seward:12} in the Large Magellanic Cloud,
both of which were estimated to be of similar age to SS433.

A number of important conclusions follow from this discovery.

If the neutron star formed through core collapse of a massive star
then the very young age of the system implies that the companion star
cannot be a low mass star, given the orbital constraints derived by
\citet{jonker:07}, because a low mass star would not have had time to
evolve off the main sequence to fill its Roche lobe at periastron.  A
massive companion star, most likely an A0 to B5 type supergiant
\citep{jonker:07}, would be consistent with the young age.

The young age of the system explains the high eccentricity and the
short orbital evolution time, between $P/\dot{P}\,\sim\,3,000$ years
inferred from X-ray dip timing \citep{parkinson:03,clarkson:04} and
$P/\dot{P}\,\sim\,20,000$ years from radio flare timing
\citep{nicolson:07}, since the system has not had time to tidally
circularize its orbit from the eccentricity it received in the
supernova explosion.

Because the system is so young, the current orbital parameters are
likely close to the orbit of the binary immediately after the
supernova explosion.  The orbital period of $P=16.5\,{\rm\,days}$ and
the eccentricity of $e\,\sim\,0.45$ \citep{jonker:07} fall into the
expected range of orbital parameters for post supernova HMXBs with
kick velocities of several hundred km/s.  In systems like this, the
neutron star spin, the companion spin, and the orbital axis are likely
mutually misaligned \citep{brandt:95}.  Spin-orbit coupling effects
can cause precession of the binary orbit and the neutron star spin
relative to each other and to the spin of the companion star,
consistent with suggestions of precession of the jet axis based on
radio \citep{calvelo:12} and X-ray data \citep{sell:10}.  Such
precession may induce changes in the accretion geometry that could
help explain the strong long term modulation of the X-ray lightcurve
\citep{brandt:95}.

While we consider a core collapse supernova the most likely
explanation for the formation of Circinus X-1, a proposed alternative
channel for neutron star formation is the accretion- induced collapse
(AIC) of a white dwarf
\citep[e.g.,][]{canal:76,nomoto:79,michel:87,bhattacharya:91}.
In this case, the companion would most likely be an evolved low mass
star (likely of mass $M \approx 0.4\,M_{\odot}$; \citealt{jonker:07}).
Given the uncertainties in the supernova parameters and the current
lack of a firm theoretical understanding of the observational
signatures of AIC supernova remnants, we cannot rule out or confirm
AIC as a formation scenario.

However, the rapid current orbital evolution of the system, with an
orbital time scale of the order of a few thousand years, and the high
orbital eccentricity, are hard to understand if the neutron star had
formed through AIC: Given that AIC supernovae are generally not
believed to impart significant eccentricity to post-supernova binary
orbits \citep{podsiadlowski:04,tauris:13}, the current orbital
parameters would reflect the pre-supernova orbit in this scenario.
The progenitor white dwarf would have had to accrete of order
0.3$M_{\odot}$ from its companion on time scales much longer than the
current orbital evolution time, and the orbit would have circularized.

At an estimated un-absorbed 0.5-10 keV surface brightness of
$\Sigma_{0.5,10} \sim 3.5^{+1.2}_{-0.8}\times 10^{-16}\,{\rm
  ergs\,s^{-1}\,cm^{-2}\,arcsec^{-2}}$ (averaged over the spectral
extraction region), the remnant has relatively low surface
  brightness compared to other remnants of similar physical size,
comparable to SN 1006, though not as faint as plerionic supernova
remnants like G21.5-0.9 \citep{bocchino:05,matheson:10}.  The low
surface brightness likely indicates that the supernova exploded into a
low density wind driven by the progenitor star
\citep{dwarkadas:05,chevalier:05}, consistent with the low ambient
Hydrogen density estimate of $n_{0} \approx
  0.27^{+0.60}_{-0.05}\,d_{8}^{-1}\,{\rm cm^{-3}}$ inferred from the
spectral fits.  In this case, the age derived from the {\tt SEDOV}
fits would likely be an overestimate \citep{dwarkadas:05}.

The estimated swept up mass is $M_{\rm s} \approx
  7.4^{+16.5}_{-1.5} \,d_{8}^{2}\,M_{\odot}$, suggesting that the
remnant has only recently entered the Sedov phase, where the swept up
mass exceeds the ejecta mass.  If the remnant is still expanding into
the stellar wind bubble blown by its progenitor, we can place a rough
lower limit on the progenitor mass of $M_{\rm prog}\,=\,M_{\rm
  ejecta}\,+\,M_{\rm wind}\,+\,M_{\rm NS}\,\gtrsim\,7\,M_{\odot}$.
The low ambient density and the roughly solar metalicity of the
emitting gas would be consistent with a type IIP supernova explosion
of a progenitor star of mass $M \sim 8$ to $25\,M_{\odot}$
\citep{chevalier:05}.

Based on the evidence that the X-ray nebula is the supernova remnant
of Circinus X-1, we conclude that the radio nebula surrounding
Circinus X-1 is most likely not inflated by the radio jets of the
source, but created through synchrotron emission from standard
non-thermal acceleration of electrons in the external shock, as first
suggested by \citet{clarkson:04}.  The limb-brightened appearance of
the radio nebula in Fig.~\ref{fig:radio} and the filamentary edge of
the nebula visible in Fig.~7 of \citet{calvelo:12b}, typical for
supernova remnants, is inconsistent with the traditional
interpretation of the nebula as a radio lobe.  While a jet-inflated
bubble of relativistic plasma could conceivably produce a similar
external shock in both non-thermal radio and thermal X-ray emission,
effectively mimicking a supernova remnant in appearance, the
synchrotron emission from within the bubble required to drive such a
powerful shock would be five orders of magnitude brighter than the
observed radio flux from the entire nebula \citep{tudose:06}, ruling
out the possibility that the large scale nebula is jet powered.

The synchrotron age of $t_{\rm jet} \sim 1,600$ years inferred for the
X-ray jets \citep{sell:10} implies that the jets, and thus accretion,
must have turned on soon after the explosion.  The location of the
synchrotron X-ray shock about a third of the way across the remnant
might indicate the location where the jets impact the inner edge of
the supernova ejecta.  The fact that both SS433 and Circinus X-1
exhibit extended synchrotron jet emission suggests that the
high-pressure environment of the interior of a supernova remnant
boosts synchrotron emission by confining the relativistic plasma at
high pressure.

The asymmetric shape of the radio nebula is more pronounced than in
other remnants of core collapse supernovae, with an aspect ratio of
more than 2:1 in the North-South vs.~East-West direction.  This may be
a result of an asymmetric explosion and/or a highly anisotropic
external medium shaped by the progenitor wind; winds from evolved
massive stars in binary systems are known to be highly anisotropic.
It is also possible that the Southern protrusion is at least in part
caused by energy injection from the jets.  This would imply a jet
power in excess of $10^{39}\,{\rm ergs\,s^{-1}}$, significantly larger
than the Eddington luminosity for a neutron star, but similar to the
jet power of SS433.  Given the similarities in the shapes of the
supernova remnants and in the properties of their relativistic jets,
Circinus X-1 may present an earlier evolutionary stage of an
SS433-like system.

The high visible extinction towards Circinus X-1 explains why the
remnant has not been found in narrow-band visible images of the
region. However, the supernova itself would have been visible to the
naked eye from latitudes lower than about 35 degrees North.

Finally, the very active ongoing accretion from the companion star,
the lack of X-ray pulses, and the observed type I X-ray thermonuclear
bursts of the source \citep{tennant:86,linares:10} imply that the
surface magnetic field of the neutron star is low, $B \ll
  10^{12}\,{\rm G}$ \citep{fujimoto:81,bildsten:98}.  Any initially
high magnetic field cannot have decayed by diffusion alone
  in the short time since the supernova
  \citep[e.g.,][]{goldreich:92,harding:06}.  Thus, the neutron star
must either have been born with low magnetic field (possibly by burial
through massive fall-back accretion;
\citealt{muslimov:95,ho:11,bernal:13}, or if the neutron star formed
through accretion-induced collapse of a low-field white dwarf;
\citealt{bhattacharya:91}) or the field must have been buried by the
rapid accretion near the Eddington rate
\citep{cumming:01,payne:06,payne:07} that the source has experienced
in the recent past.  Given the low magnetic field strength
  and the implied low rate of magnetic braking, the neutron star is
  likely spinning at close to the rotation rate it had at birth.

The discovery of the supernova remnant of Circinus X-1 opens a new
window on the study of young X-ray binaries.  The ability to correlate
the rapid, non-linear evolution of the orbital parameters with the
accretion behavior of the source holds significant promise for a
detailed understanding of the complex, sometimes puzzling behavior of
a newly formed X-ray binary.

\acknowledgements We would like to thank Jay Gallagher, Ellen Zweibel,
and Snezana Stanimirovic for helpful discussions.  S.H.~and
P.S.~acknowledge support through CXC grant G09-0056X and NSF grant
AST-0908690; W.N.B.~acknowledges support through NASA ADP grant
NNX10AC99G and CXC contract SV4-74018. This research has made use of
data obtained from the Chandra Data Archive and the Chandra Source
Catalog, and software provided by the Chandra X-ray Center (CXC) in
the application packages included in CIAO.  The Australia Telescope
Compact Array is part of the Australia Telescope National Facility
which is funded by the Commonwealth of Australia for operation as a
National Facility managed by CSIRO.

\end{document}